\documentclass{revtex4-1}  

\usepackage{amsmath,amsthm,graphicx}
\usepackage{dsfont}
\usepackage{graphics}
\usepackage{color}
\begin{document}
 
\newcommand{\bra}[1]{\langle #1|}

\newcommand{\ket}[1]{|#1\rangle}

\newcommand{\braket}[2]{\langle #1|#2\rangle}

\title{Optimal measurement strategies for the trine states with arbitrary prior probabilities}
\author{Graeme Weir}
\address{School of Physics and Astronomy, University of Glasgow, Glasgow G12 8QQ, United Kingdom}
\author{Catherine Hughes}
\address{School of Physics and Astronomy, University of Glasgow, Glasgow G12 8QQ, United Kingdom}
\author{Stephen M. Barnett}
\address{School of Physics and Astronomy, University of Glasgow, Glasgow G12 8QQ, United Kingdom}
\author{Sarah Croke}
\address{School of Physics and Astronomy, University of Glasgow, Glasgow G12 8QQ, United Kingdom}

\begin{abstract}
We investigate the optimal measurement strategy for state discrimination of the trine ensemble of qubit states prepared with arbitrary prior probabilities. Our approach generates the minimum achievable probability of error and also the maximum confidence strategy. Although various cases with symmetry have been considered and solution techniques put forward in the literature, to our knowledge this is only the second such closed form, analytical, arbitrary prior, example available for the minimum-error figure of merit, after the simplest and well-known two-state example.
\end{abstract}

\maketitle

\section{Introduction}

Quantum key distribution, and quantum communication in general, depends on the problem of quantum state discrimination \cite{bae2015quantum,barnettcroke2009quantum}. The standard formulation of this problem involves two communicating parties, Alice and Bob: Alice communicates with Bob by sending him a quantum state $\rho_i$ which has been chosen from a set of possible states $\{\rho_j\}$, each with an \emph{a priori} probability $p_j$. Bob knows these states and their probabilities, and his goal is to determine which state was sent, thereby decoding the message which Alice wishes to communicate. Clearly, Bob wishes to decode the message as best he can, and he may quantify this using any of a number of different figures of merit. The two most common figures of merit he might wish to maximise are mutual information \cite{davies1978information,sasaki1999accessible,levitin1994entropy}, and the probability of correctly identifying the state \cite{helstrom1976quantum,yuen1975optimum,bae2013minimum,ha2013complete}. He may also use the techniques of unambiguous discrimination \cite{chefles1998unambiguous,chefles2001unambiguous} -- which either gives an inconclusive outcome or identifies the signal state with certainty -- or maximum confidence, a generalisation of this which sometimes yields incorrect answers \cite{croke2006maximum}.

This has been a popular problem for a few decades, with theoretical solutions obtained and experiments performed for various sets of states and figures of merit \cite{sasaki1999accessible,bae2013minimum,ha2013complete,chefles1998unambiguous,chefles2001unambiguous,croke2006maximum,ivanovic1987differentiate,dieks1988overlap,peres1988differentiate,jaeger1995optimal,mosley2006experimental,chitambar2013revisiting,andersson2002minimum,ban1997optimum,barnett2001minimum,chou2004minimum,clarke2001experimental,clarke2001experimentalUSD,mohseni2004optical,huttner1996unambiguous,weir2017optimal}. This popularity may be attributed partially to the fundamental nature of the problem, and also to its far-reaching consequences: in addition to being crucial for quantum key distribution, state discrimination has relevance in quantum information processing and quantum metrology \cite{bae2015quantum}, and also allows us to explore the constraints on different measurement classes such as global measurement or local measurement with classical feed-forward \cite{walgate2000local, bennett1999quantum, crokeweir2017}. For minimum error and unambiguous discrimination, the problem of optimisation may be cast as a semi-definite programme, and for particular instances efficient algorithms exist \cite{jevzek2002finding}. Explicit analytic solutions are available only for the simplest cases, however, despite recent progress in analytical techniques for minimum error discrimination.

The optimal measurement for discriminating between two qubit states with arbitrary prior probabilities is known for both the minimum error \cite{barnettcroke2009quantum} and maximum mutual information figures of merit, and indeed the strategies coincide in the two-state case \cite{levitin1994entropy}. For two pure qubit states the optimal unambiguous discrimination measurement is also known for arbitrary priors \cite{barnettcroke2009quantum,helstrom1976quantum,jaeger1995optimal}.

The trine ensemble, consisting of three pure states spaced symmetrically on the Bloch sphere, is the next simplest case, and the simplest for which a full analytic solution for arbitrary priors is not available for any of the most commonly used strategies. An analytic solution is desirable for a number of reasons: it allows us to understand, qualitatively and quantitatively, how the optimal measurements and corresponding figures of merit depend on the prior probabilities; it may be used in a larger optimisation problem in which state discrimination arises as just one part; and it allows comparisons to the simpler two-state example which is well-known. Indeed, it is already known that there are qualitative differences between the two- and three-state examples: for the trine states with equal priors, ($p_i=\frac{1}{3}$), the strategy for minimum-error discrimination \cite{ban1997optimum} is different than that for maximising the mutual information \cite{sasaki1999accessible}, in contrast to the two-state example.

Further, in the case of three qubit states, the minimum-error measurement has been found analytically for a generalised version of the trine states with symmetric prior probabilities, i.e. $p_0=1-2p, p_1=p_2=p$ \cite{andersson2002minimum}, with the interesting result that the number of measurement outcomes is dependent on the specific parameters chosen for the set of states.

In this paper, we give a complete analysis of the problem of state discrimination for the trine states with arbitrary prior probabilities, for both the minimum error and maximum confidence figures of merit. Each of these are amenable to analytic solutions; in the minimum error case, which we begin with, this is made possible by recent developments \cite{ha2013complete,weir2017optimal,deconinck2010qubit}. We continue by investigating the maximum confidence measurement \cite{croke2006maximum} for the trine states with arbitrary prior probabilities and obtain an expression for the probability of correctly identifying each signal state using this measurement scheme.

\section{Minimum Error Measurements}\label{trine}

We begin by reviewing the description of measurement in quantum theory, which we will use throughout the paper. Any physically allowed measurement is described mathematically by a POVM (positive-operator valued measure), consisting of a set of Hermitian operators \{$\pi_i$\}, individually called POVM elements, which satisfy the following conditions:

\begin{align}
\nonumber \pi_i&\geq 0\\
\nonumber \sum_i \pi_i &=\mathds{1}.
\end{align}
The Born rule tells us the probability of obtaining any outcome $j$ -- corresponding to a ``click" at the detector associated with POVM element $\pi_j$ -- when measuring a system prepared in state $\rho$:

\begin{equation}\label{Born}
P(j|\rho)=\rm{Tr}(\rho\pi_j).
\end{equation}
The above conditions on the POVM elements simply ensure that the probabilities given by the Born rule are all positive and sum to 1.

\begin{figure}
\centering
\includegraphics[scale=0.3]{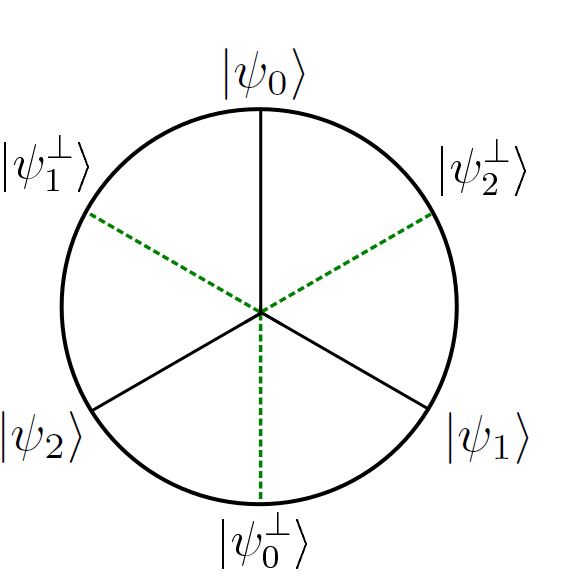}
\caption{The trine states on the equator of the Bloch sphere. The dotted lines show the anti-trine measurement basis - each POVM element is aligned so that it is orthogonal to one of the potential states. For instance, if we get a ``click" at the POVM element at $\ket{\psi_0^\perp}$, we know with certainty that state $\ket{\psi_0}$ was not prepared.}
\label{AntiTrine}
\end{figure}

In this paper we are concerned with the trine states, qubit states associated with three equidistant points on any great circle of the Bloch sphere. We will place the trine states on the equator of the Bloch sphere, so that:

\begin{align*}
\ket{\psi_0}&=\frac{1}{\sqrt{2}}(\ket{0}+\ket{1}),\\
	\ket{\psi_1}&=\frac{1}{\sqrt{2}}(\ket{0}+e^{i\frac{2\pi}{3}}\ket{1}),\\
\ket{\psi_2}&=\frac{1}{\sqrt{2}}(\ket{0}+e^{i\frac{4\pi}{3}}\ket{1}),
\end{align*}
where the states $\ket{0}$ and $\ket{1}$ correspond to the poles on the Bloch sphere. These trine states can be visualised on the Bloch sphere as shown in figure \ref{AntiTrine}. For equal prior probabilities ($p_0=p_1=p_2=\frac{1}{3}$), it is known that the optimal measurement of the trine states for minimising the probability of error is to measure along the states themselves \cite{ban1997optimum}, that is, making a measurement of the form $\pi_j=\frac{2}{3}\ket{\psi_j}\bra{\psi_j}$. This is known as the trine measurement.

In contrast to a two-state system, intriguingly, if we wish to maximise the mutual information gained by our measurement, we must use a different POVM: in this case, we perform the so-called anti-trine measurement \cite{clarke2001experimental}, as shown in figure \ref{AntiTrine}. This involves three measurement outcomes, each of which is perpendicular to one of the trine states; this is therefore an eliminatory measurement, as it tells us with certainty that the system was \emph{not} prepared in a particular state, with the other two possible states equally likely to be the signal state.

Throughout this paper, and without loss of generality, we assume $p_0\geq p_1 \geq p_2$.

\subsection{Helstrom conditions}
A POVM \{$\pi_i$\} is optimal for minimum-error discrimination between a set of states \{$\rho_i$\} with prior probabilities \{$p_i$\} if and only if it satisfies the Helstrom conditions \cite{helstrom1976quantum,barnett2009quantum}:

\begin{align}
&\pi_i(p_i \rho_i-p_j\rho_j)\pi_j=0 \quad \forall i,j \label{H1}\\
&\sum_{i}p_i\rho_i\pi_i-p_j\rho_j\geq 0 \quad \forall j, \label{H2}
\end{align}
where outcome $i$, corresponding to element $\pi_i$, is taken to indicate that the state $\rho_i$ was transmitted. Note that the probability of Bob correctly guessing the signal state is given by Born's Rule, in equation (\ref{Born}):

\begin{equation}
P_{\rm{Corr}}=\sum_{i=0}^{n-1}p_i\rm{Tr}(\rho_i\pi_i)
\end{equation}
The minimum-error measurement scheme is defined as one which maximises the above quantity.

In the case of the trine states, the minimum-error measurement must have either two or three elements: a one-element measurement, that is $\pi_k=\mathds{1}$ for some $k$, corresponding to the ``no-measurement" strategy, can never be optimal for pure state ensembles \cite{hunter2003measurement}, as condition (\ref{H2}) cannot be satisfied for $j\neq k$. Furthermore, as each measurement outcome corresponds to identifying one of the potential states, the number of outcomes cannot exceed the number of states: any extra elements will be redundant.


In light of this, we split the problem into two parts: we ask when a two-element POVM is optimal, as this is a relatively easy problem to solve, and then we consider the remaining parameter space. In the region where the two-outcome measurement does not give the minimum error, we know that a three-element POVM of some form will be optimal. In this region, we construct the optimal measurement by applying the strategy outlined in \cite{weir2017optimal}. Surprisingly, the two-element POVM is optimal for almost the whole parameter space. We show, explicitly, that all optimal measurements on the trine states are unique - that is, for any choice of initial probabilities \{$p_i$\}, there is one and only one measurement which is optimal.

\subsection{Conditions for a two-element POVM to be optimal}\label{2ElOpt}

\begin{figure}
\centering
\includegraphics[scale=0.12]{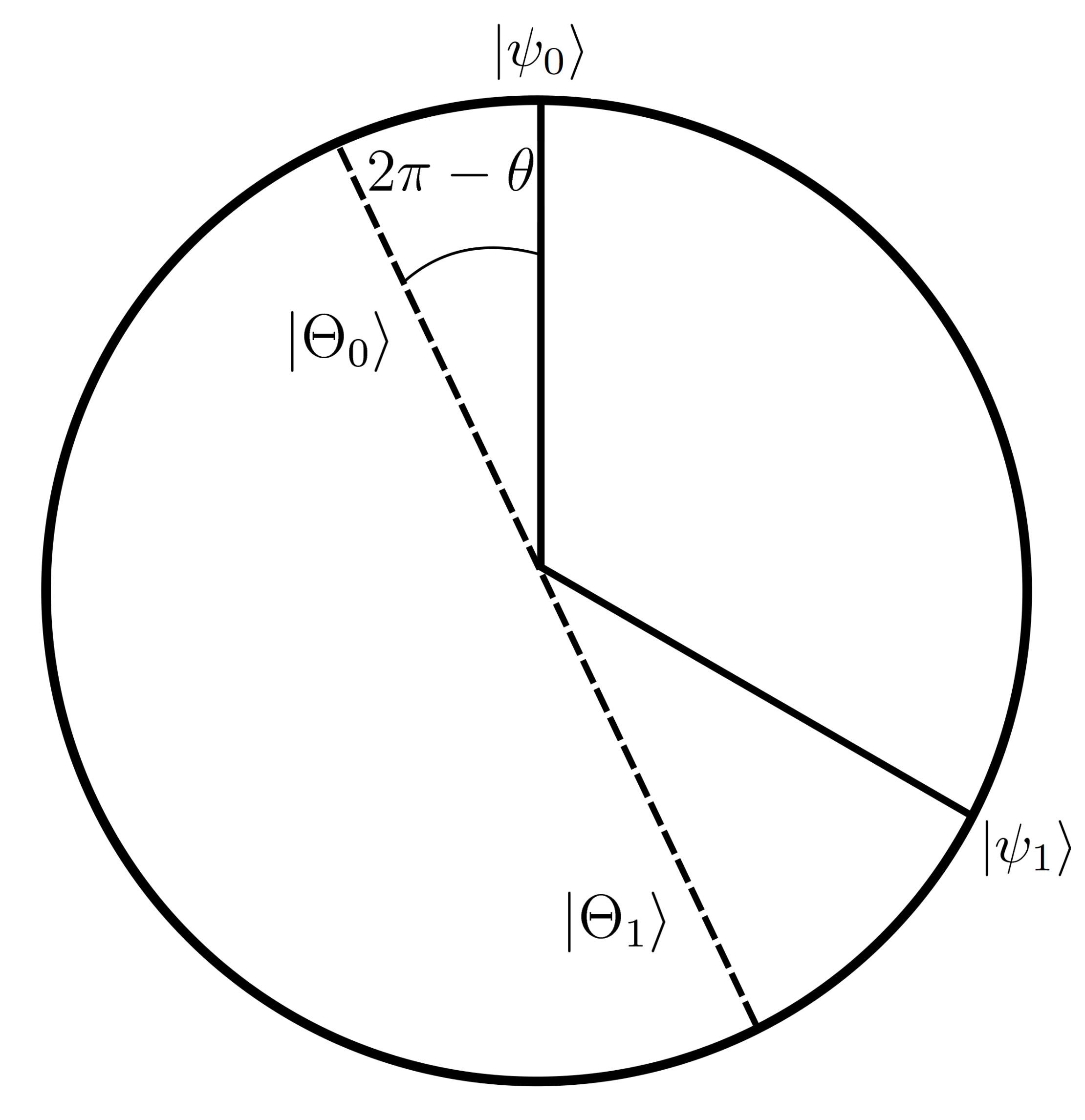}
\caption{The two signal states we are trying to discriminate between ($\ket{\psi_0}$ and $\ket{\psi_1}$, solid lines) and the optimal measurement for doing so (dotted lines), where $\theta$ is as defined in equation (\ref{tan}). Note that the signal states are not symmetrical with respect to the measurement states - the optimal measurement is biased towards identifying $\ket{\psi_0}$, the \emph{a priori} more likely state.}
\label{TanMment}
\end{figure}

We know that when $p_2=0$, a two-element POVM must be optimal. This problem has a well-known solution, with the optimal probability of correctness given by the Helstrom bound \cite{barnettcroke2009quantum,helstrom1976quantum}:

\begin{equation}
P_{2-el}=\frac{1}{2}(1+\sqrt{1-4p_0p_1|\braket{\psi_0}{\psi_1}|^2}),
\end{equation}
where ``2-el" is short for two-element. It is readily shown that this is achieved by a measurement of the form $\pi_{0,1}=\ket{\Theta_{0,1}}\bra{\Theta_{0,1}}$, $\pi_2=0$, where

\begin{align*}
\ket{\Theta_0}&=\frac{1}{\sqrt{2}}(\ket{0}+e^{i\theta}\ket{1}),\\
\ket{\Theta_1}&=\frac{1}{\sqrt{2}}(\ket{0}-e^{i\theta}\ket{1}),\\
\end{align*}
with
\begin{equation}\label{tan}
\tan\theta=\frac{-\sqrt{3} p_1}{2p_0+p_1}.
\end{equation}
Figure \ref{TanMment} shows the measurement states on the Bloch sphere. Writing the probability of correctly guessing the state in terms of only $p_0$ and $p_1$ gives:

\begin{equation}\label{OPC2}
P_{2-el}=\frac{1}{2}(p_0+p_1+\sqrt{p_0^2+p_0p_1+p_1^2})
\end{equation}

As described in \cite{ha2013complete,weir2017optimal,deconinck2010qubit}, we know that if state $\rho_2$ is added to this ensemble with a small enough probability, the number of POVM elements necessary for minimum-error measurement remains unchanged. Intuitively, if $p_2$ is small enough, we do not gain anything by identifying $\rho_2$, and the minimum-error measurement favours the more likely states. We can use the Helstrom conditions to define precisely what ``small enough" means in this context, and put conditions on $p_0, p_1$ and $p_2$ which state when a two-element POVM is sufficient and when a three-element POVM is required.

To find the values for $p_0$ and $p_1$ for which a two-element POVM is the optimal measurement, we investigate the other Helstrom condition, shown in equation (\ref{H2}). This is trivial for $j=0, 1$, as we already know this must be the optimal measurement when these are the only signal states. Therefore, it suffices to check the positivity of the matrix

\begin{equation} \label{MatrixM}
M=\sum_ip_i\rho_i\pi_i-p_2\rho_2.
\end{equation}
It follows from the conditions for $j=0,1$ that $\sum_ip_i\rho_i\pi_i$ is a positive operator. Further, as $\rho_2$ is a pure state, $M$ has at most one negative eigenvalue, and to check positivity of $M$ we can calculate the sign of the determinant: when $\det(M)$ is positive, the two-element POVM described above is optimal. This is straightforward, and the determinant of the matrix is found to be

\begin{equation}\label{SymmPoly}
\begin{split}
\det(M)=&-3 p_0^4 - 3 p_1^4 - 10 p_0^3 p_1 - 10 p_0 p_1^3+ 6 p_0^3 + 6 p_1^3 \\&- 13 p_0^2 p_1^2 + 12 p_0^2 p_1+ 12 p_0 p_1^2 - 3 p_0^2 - 3 p_1^2 - 2 p_0 p_1.
\end{split}
\end{equation}
To find the boundary of the region where the two-element measurement is optimal, it is useful to parameterise the probabilities as follows: $p_0=p+\delta, p_1=p-\delta, p_2=1-2p$, where the ordering $p_0\geq p_1 \geq p_2$ implies $\delta \geq 0, \delta \leq 3p-1, \delta \leq p$.

\begin{figure}[!ht]
	\centering
	\includegraphics[width=0.47\columnwidth]{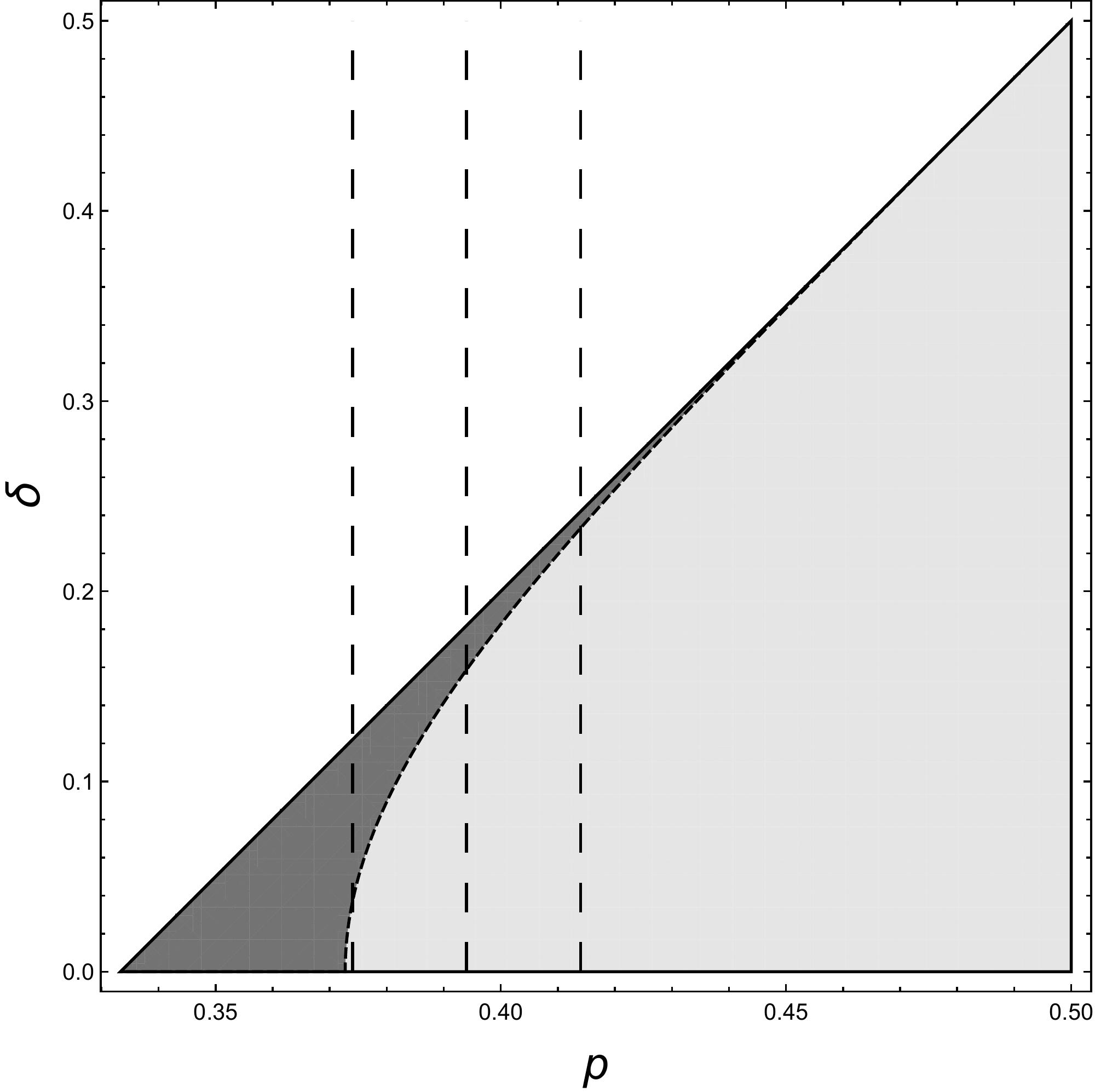}
	\caption{Graph showing the sign of the determinant of matrix $M$ in equation (\ref{MatrixM}) as a function of $p$ and $\delta$. The dark region corresponds to a negative determinant, and hence shows the region where a 3-element POVM is optimal. The light area displays the rest of the allowable parameter space, where the 2-element POVM we have discussed is optimal. The three dashed vertical lines correspond to the three plots shown in figure \ref{Threeplots}. Note that the diagonal line $\delta=3p-1$ corresponds to $p_1=p_2$.
		\label{DeltaGraph}}
\end{figure}

After a little algebra, we find that the determinant is simply a quadratic in $\delta^2$, with roots $\pm\delta_{c\pm}$, where
\begin{equation}
\delta_{c\pm}^2=2-6p+5p^2\pm 2\sqrt{1-6p+16p^2-24p^3+16p^4}
\end{equation}
There are four roots for $\delta$, only two of which give physically-realisable probability distributions; these two simply swap $p_0$ for $p_1$ and vice-versa (the other two roots correspond to unphysical distributions with, e.g., $p_0>1$).
Imposing our condition that $p_0\geq p_1 \geq p_2$, we find that $\det(M)\geq 0$ for $\delta\leq\delta_{c-}$. That is, a two-element POVM is optimal when $\delta < (2-6p+5p^2 - 2\sqrt{1-6p+16p^2-24p^3+16p^4})^{\frac{1}{2}}$. Otherwise some three-element POVM (discussed in the next section) is optimal. The parameter regions for which the optimal measurement has two or three outcomes are shown in figure \ref{DeltaGraph}.

It is apparent that a three-element POVM is only optimal when close to a symmetric ensemble, i.e. $p_1$ very close to $p_2$. For $p_1=p_0 \in [\frac{1}{3}, \frac{4}{9+\sqrt{3}})$ and for all $p_1=p_2$, the symmetric three-element measurement outlined in \cite{andersson2002minimum} is optimal.

An interesting consequence of equation (\ref{tan}) is that there is not a one-to-one correspondence between ensembles and optimal measurements. As we can increase $p_2$ from zero without changing the optimal measurement, there are many different ensembles with the same optimal measurement strategy. In this region, where the two-element POVM is optimal, the optimal measurement depends only on the relative frequency of occurrence of $p_0$ and $p_1$ (i.e. the ratio between $p_0$ and $p_1$). For fixed measurement angle $\phi$, the probability of correctness increases linearly with $p_0+p_1$. We also note that this effect does not happen in the two-state discrimination case, where, given two states and a measurement which is known to be optimal, there is only one $p_0$ -- and hence only one complementary $p_1$ -- which will satisfy the Helstrom conditions.


\subsection{Optimal Three-element POVM}\label{3ElOpt}

We now turn our attention to the region in which we know a three-element POVM must be optimal. This region is hard to analyse due to its lack of symmetry, but the problem can be solved analytically by using the Helstrom conditions constructively, following the approach developed in \cite{weir2017optimal}. We briefly review the method before applying it to the case of interest here.

If we take equation (\ref{H1}) and sum over $i$, using the stipulation that $\sum_i \pi_i = \mathds{1}$, we obtain

\begin{equation}
(\Gamma-p_j\rho_j)\pi_j=0 \quad \forall j,
\end{equation}
where $\Gamma=\sum_i p_i \rho_i \pi_i=\sum_i p_i \pi_i \rho_i$. This is a $2 \times 2$ matrix which, by (\ref{H2}), has non-negative determinant, which means that $\pi_j$ is either the zero matrix (which we are forbidding as we already know when a two-element POVM is optimal) or is perpendicular to $\Gamma-p_j\rho_j$. That is, if $\pi_j=k_j\ket{\phi_j}\bra{\phi_j}$ for some constant $k_j$, we have

\begin{equation}\label{perp}
\Gamma-p_j\rho_j=c_j\ket{\phi_j^\perp}\bra{\phi_j^\perp}
\end{equation}
for some constant $c_j$, where $\braket{\phi_j}{\phi_j^\perp}=0$. This proves that $\Gamma$ has a linearly independent decomposition consisting of the vectors $\ket{\phi_j^\perp}$ and $\ket{\psi_j}$. As such, a result from \cite{nielsen2010quantum}, also used in \cite{mochon2006family}, is applicable. It may be shown \footnote{To see this, note that equation (\ref{perp}) implies that $\mathds{1}=\Gamma^{-1}p_j\ket{\psi_j}\bra{\psi_j}+\Gamma^{-1}c_j\ket{\phi_j^{\perp}}\bra{\phi_j^{\perp}}$, and taking the matrix element $\bra{\psi_j}\cdot\ket{\phi_j}$ gives the desired result.} \cite{weir2017optimal} that:

\begin{equation}
\bra{\psi_j}\Gamma^{-1}\ket{\psi_j}=\frac{1}{p_j}.
\end{equation}

By writing $\Gamma^{-1}$ in the form $\frac{1}{2}(a\mathds{1}+\vec{b}\cdot\hat{\sigma})$, we find three linear equations in three unknowns. As described in \cite{weir2017optimal} and \cite{herzog2015optimal}, we may assume from symmetry  that the optimal POVM will be in the same plane as the states, so $b_z=0$, and hence find $a, b_x, b_y$. Thus we can find $\Gamma$ and hence $P_{\rm{Corr}}$, the optimal probability of correctly identifying the state which was sent, as $P_{\rm{Corr}}=\sum_k p_k \rm{Tr}(\rho_k\pi_k)=\rm{Tr}(\Gamma)=\frac{4a}{a^2-|b|^2}$. In fact, because we know that $\Gamma-p_j\rho_j=c_j\ket{\phi_j^\perp}\bra{\phi_j^\perp}$, we can also explicitly find the POVM elements and hence extract the optimal measurement directly from the Helstrom conditions. Furthermore, as $\Gamma$ is known to be unique for a given set of states \cite{bae2013minimum}, this POVM will be unique for this ensemble of $\{p_j\}$ and $\{\rho_j\}$, as the vector solution $\ket{\phi^{\perp}_j}$ is unique.

It is sufficient for our purposes to simply calculate $P_{3-el}$. From the above, we obtain

\begin{align*}
a&=\frac{2}{3}\bigg(\frac{1}{p_0}+\frac{1}{p_1}+\frac{1}{p_2}\bigg)\\
b_x&=\frac{2}{3}\bigg(\frac{2}{p_0}-\frac{1}{p_1}-\frac{1}{p_2}\bigg)\\
b_y&=\frac{2}{\sqrt{3}}\bigg(\frac{1}{p_1}-\frac{1}{p_2}\bigg),
\end{align*}
which yields:

\begin{equation}\label{OPC3}
P_{3-el}=\frac{2(p_0p_1+p_0p_2+p_1p_2)}{2-(\frac{p_0p_1}{p_2}+\frac{p_0p_2}{p_1}+\frac{p_1p_2}{p_0})}.
\end{equation}
If we compare this to the expression for $P_{2-el}$ given by the optimal two-element POVM then we find that they meet at the boundary when the two-element POVM stops being optimal, as we would expect.

Our expression for $P_{3-el}$ has the interesting property that, in parts of the region where we know the two-element POVM to be optimal, the expression for $P_{3-el}$ yields a greater value than $P_{2-el}$. We also obtain some values for $P_{3-el}$ which are greater than 1, which is clearly incorrect. These anomalies are due to the fact that our method of obtaining $\Gamma$ does not \emph{strictly} impose the conditions for POVM elements; specifically, not every POVM element $\pi_i$ is a positive semi-definite operator. This may be seen by comparing our measurement to the analogous measurement detailed in \cite{andersson2002minimum}. This is not a problem, of course, as these regions in the parameter space are readily determined. At $\delta=0$, we have $p_0=p_1$ and the optimal measurement includes a POVM element of the form $(1-a^2)\ket{\psi_2}\bra{\psi_2}$, with $a=\frac{\sqrt{3}p}{4-9p}$. Clearly the factor of $(1-a^2)$ becomes negative for $p>\frac{4}{9+\sqrt{3}}$, and so our attempted measurement no longer fulfils the POVM criteria, giving spurious results. It is at this point that the two-element POVM becomes optimal. Thus we can conclude that our optimal three-element POVM does indeed become invalid in the region where we know a two-element POVM must be optimal.

\begin{figure}[!hb]
	\centering
	\includegraphics[width=0.47\columnwidth]{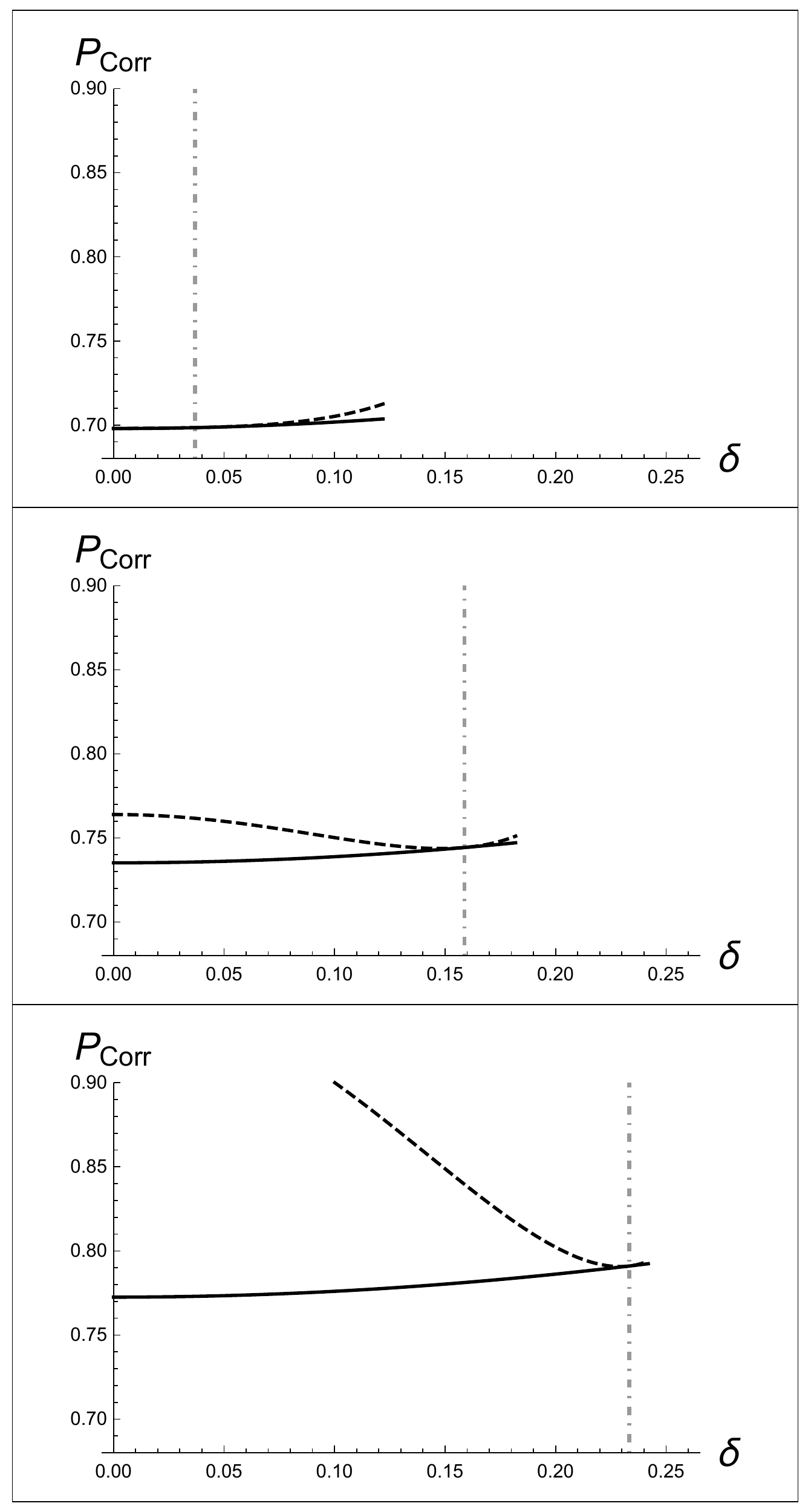}
	\caption{Comparisons of $P_{\rm{Corr}}$ given by the optimal two-element POVM (bold line) and the results given by our method for finding the optimal three-element POVM (dotted line) for fixed values of $p$. From top to bottom, $p$ has values $0.374$, $0.394$, and $0.414$. The dot-dashed grey vertical lines show when the determinant in equation (\ref{SymmPoly}) becomes negative and thus a three-element POVM becomes physically realisable. That is, the three-element POVM is only viable to the right of the dot-dashed grey line. Note that, when physically viable, the three-element POVM does not significantly outperform the two-element POVM.
		\label{Threeplots}}
\end{figure}

\begin{figure}[!ht]
\centering
\includegraphics[width=0.5\columnwidth]{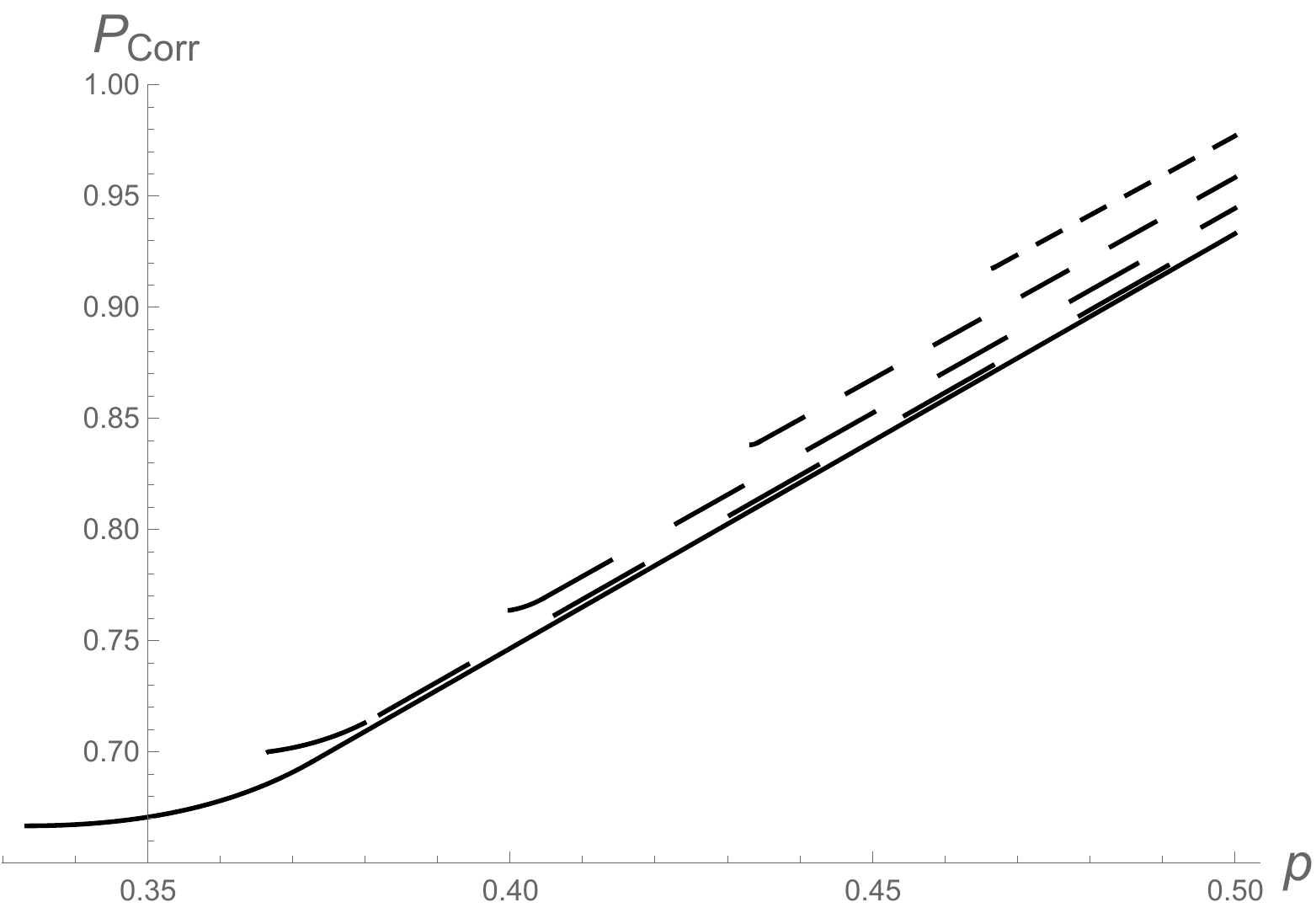}
\caption{Graph showing the probability of correctly identifying the signal state using the optimal measurement strategy for $p\in[\frac{1}{3}, \frac{1}{2}]$. The lines, in increasing amounts of dashing - and lowest to highest - correspond to $\delta=0, \delta=0.1, \delta=0.2, \delta=0.3$ and $\delta=0.4$.}
\label{PCorr}
\end{figure}

To summarise, this gives us the following functions for the probability of correctly guessing the signal state using the minimum-error measurement scheme. In the case where $\delta < (2-6p+5p^2$ $- 2\sqrt{1-6p+16p^2-24p^3+16p^4})^{\frac{1}{2}}$, we have:

\begin{align}\label{P2el}
P_{2-el}&=\frac{1}{2}(p_0+p_1+\sqrt{p_0^2+p_0p_1+p_1^2})\\
&\nonumber=p+\frac{1}{2}\sqrt{3p^2+\delta^2}.
\end{align}
Otherwise:
\begin{align}\label{P3el}
P_{3-el}&=\frac{2(p_0p_1+p_0p_2+p_1p_2)}{2-(\frac{p_0p_1}{p_2}+\frac{p_0p_2}{p_1}+\frac{p_1p_2}{p_0})}\\
&\nonumber=\frac{2(1-2p)(p^2-\delta^2)(3p^2+\delta^2-2p)}{9p^4-4p^3+6p^2\delta^2-12p\delta^2+4\delta^2+\delta^4}.
\end{align}
We can therefore plot the optimal probability of correctness for discriminating between the trine states for arbitrary prior probabilities. These results are shown in figure \ref{Threeplots} and figure \ref{PCorr}, for various values of $p$ and $\delta$.

This solves the problem of minimum-error state discrimination between the trine states for \emph{all possible} probability distributions, and highlights some differences between two-state and three-state discrimination. Firstly, for the two-state case we always require two POVM elements and, indeed, these are both simple projectors. In this case each signal state has a measurement outcome associated with it. This is not the case for the three-state problem, for which it is sometimes beneficial to simply never measure one of the signal states. Indeed, for most of the parameter space, a two-outcome measurement is optimal. Our solution also shows that, for three states, there is not a one-to-one correspondence between ensembles and optimal measurements; a certain measurement strategy may be optimal for multiple probability distributions of the trine states, whereas in the two-state case each optimal measurement strategy is unique to its corresponding probability distribution.

For $\delta=0$ our results agree with previous work \cite{andersson2002minimum,weir2017optimal}. As we have produced an analytic solution, it is also possible to use this to solve problems where state discrimination arises as a smaller part of a problem, as occurs when multiple copies are available \cite{crokeweir2017}.

\section{Maximum Confidence Measurements}

Maximum confidence measurements may be viewed as a generalisation of unambiguous discrimination \cite{croke2006maximum,mosley2006experimental}: whereas the latter is only possible when the states to be measured are linearly independent \cite{chefles1998unambiguous}, maximum confidence is a viable strategy for linearly dependent states. While the maximum confidence measurement does not have the advantage of giving an answer which is \emph{guaranteed} to be correct (as unambiguous discrimination does), it offers a ``middle ground" where, if a given state is identified, it is with the lowest possible probability of error for that state; otherwise the output is an inconclusive outcome, similarly to unambiguous discrimination. It has the advantage of an analytic solution for the elements of the optimal POVM in general, and is also related in certain cases to the minimum-error strategy. Understanding the maximum confidence measurement for the trines with arbitrary priors provides insight into the form of our minimum-error results.

The maximum confidence measurement scheme has already been described for three equiprobable symmetric states on the Bloch sphere \cite{croke2006maximum,croke2007maximum}, and we extend this to the case with arbitrary prior probabilities.

In this measurement scheme, we have $\pi_i \propto \rho^{-1} \rho_i \rho^{-1}$, where $\rho=\sum_j p_j \ket{\psi_j}\bra{\psi_j}$. Note that the figure of merit for this strategy is the probability of outcome $\pi_i$ correctly identifying the state $\rho_i$, given by Bayes:

\begin{equation}\label{Confidence}
P(\rho_i|\pi_i)=\frac{p_iP(\pi_i|\rho_i)}{P(\pi_i)}=\frac{P(\pi_i,\rho_i)}{P(\pi_i,\rho_i)+\sum_{j\neq i}P(\pi_i,\rho_j)}.
\end{equation}
This is independent of the constant of proportionality multiplying $\pi_i$, which may therefore be chosen arbitrarily. It is always possible to choose the constants of proportionality such that $\sum_j\pi_j\leq\mathds{1}$. If necessary, a complete measurement may then be formed by adding an inconclusive outcome $\pi_?=\mathds{1}-\sum_j\pi_j$. The probability that each measurement outcome accurately reflects the state of the system is, however, independent of how we choose to complete the measurement.

It is convenient to note that, for qubits, $\rho^{-1}\propto\sum_j p_j \ket{\psi_j^{\perp}}\bra{\psi_j^{\perp}}$. This may be seen by considering $\rho$ as a point within the Bloch sphere; $\rho^{-1}$ must therefore correspond to the antipodal point in the Bloch sphere - this antipodal point is given by $\sum_j p_j \ket{\psi_j^{\perp}}\bra{\psi_j^{\perp}}$. It is perhaps useful to think of the decompositions of $\rho=\frac{1}{2}(\mathds{1}+\vec{b}\cdot \hat{\sigma})$ and $\rho^{-1}\propto(\mathds{1}-\vec{b}\cdot \hat{\sigma})$. In fact, we can go further by noting that $\rho^{-1}=[1-\rm{Tr}(\rho^2)]^{-1}(\mathds{1}-\vec{b}\cdot\hat{\sigma})$: this means $\rho$ must be a mixed state, otherwise $\rho^{-1}$ has no physical meaning.
Using $\rho^{-1}\propto\sum_j p_j \ket{\psi_j^{\perp}}\bra{\psi_j^{\perp}}$, therefore, we may write
\begin{equation}
\pi_i \propto \sum_{j, k}p_jp_k\ket{\psi_j^\perp}\braket{\psi_j^\perp}{\psi_i}\braket{\psi_i}{\psi_k^\perp}\bra{\psi_k^\perp}.
\end{equation}

The numerator of equation (\ref{Confidence}) in the general case is $p_i\bra{\psi_i}\pi_i\ket{\psi_i}\propto p_i(\sum_mp_m|\braket{\psi_m^\perp}{\psi_i}|^2)^2$. Due to the symmetry of the trine ensemble, it is readily verified that $|\braket{\psi_j^\perp}{\psi_i}|^2=\frac{3}{4}(1-\delta_{ij})$. The numerator, in this instance, is therefore $\frac{9}{16}p_i(1-p_i)^2$.
The other piece of this expression takes the following form, where the last two lines are dependent on the number of states we are discriminating between and their overlaps:

\begin{align*}
\sum_{j\neq i}P(\pi_i, \rho_j)&= \sum_{j\neq i}p_j\bra{\psi_j}\pi_i\ket{\psi_j}\\
&\propto \sum_{j\neq i} p_j \left|\sum_m p_m \braket{\psi_j}{\psi_m^\perp}\braket{\psi_m^\perp}{\psi_i}\right|^2\\
&= \frac{9}{16}\sum_{j\neq i} p_j \sum_{m\neq i, j}p_m^2\\
&= \frac{9}{16}(1-p_i)\prod_{j\neq i} p_j\\
\end{align*}
The final line may not be obvious at first, but can be verified by setting, e.g., $i=0$ and noting that $m$ can only take one value - if $j=1$, $m=2$ and vice versa. We therefore obtain 

\begin{equation}
P(i)_{\rm{Corr}}=\left(1+\frac{\prod_{j\neq i} p_j}{p_i\sum_{j\neq i}p_j}\right)^{-1},
\end{equation}
which has some attributes we might expect: when any individual $p_j$ is set equal to zero, the probability of correctly identifying the state $\rho_i$ ($i\neq j$) becomes unity, as the set of possible states is now linearly independent, allowing unambiguous discrimination to be performed. When $p_i$ is zero, there is zero chance of that state being correctly identified, as one might anticipate.

\begin{figure}[!ht]
	\centering
	\includegraphics[width=0.47\columnwidth]{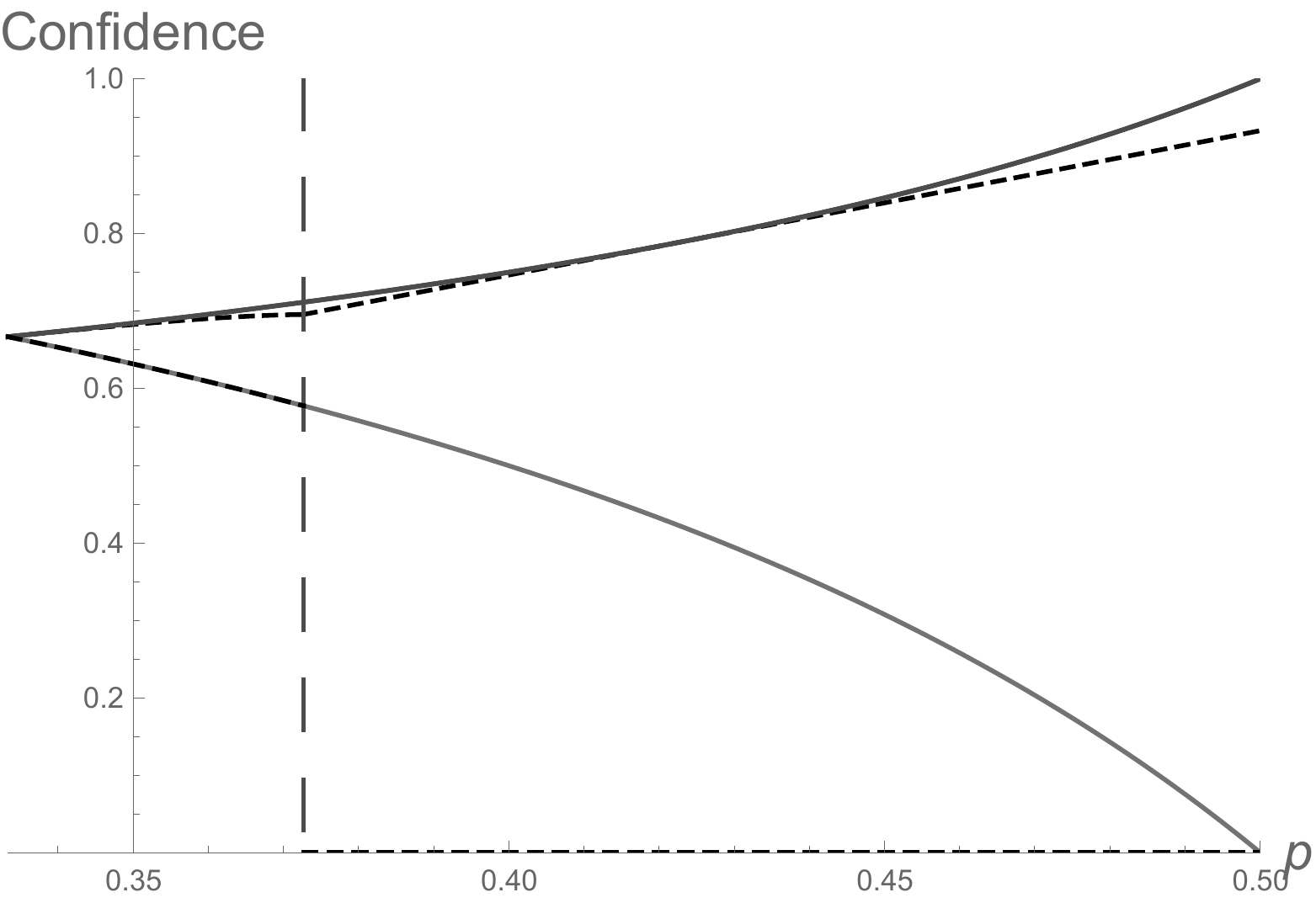}
	\caption{Graph showing the confidence in correctly identifying the signal state given the outcome of the minimum-error strategy (black dotted line) and the maximum confidence measurement for $p\in[\frac{1}{3}, \frac{1}{2}]$ and $\delta=0$. The lighter lines (from darkest to lightest and top to bottom) represent the MCM strategy on states $\rho_0, \rho_1$ and $\rho_2$ - note that as $\delta=0$, the states $\rho_0$ and $\rho_1$ are equally likely, and so their values for confidence completely overlap. Also, the minimum-error measurement and maximum confidence measurement are identical for $\rho_2$, so give the same confidence value, resulting in only $3$ lines being visible. The dotted vertical line corresponds to the crossover point at which the minimum-error measurement stops being a three-outcome measurement and starts being a two-outcome measurement.
		\label{MCM1}}
\end{figure}

\begin{figure}[!ht]
	\centering
	\includegraphics[width=0.47\columnwidth]{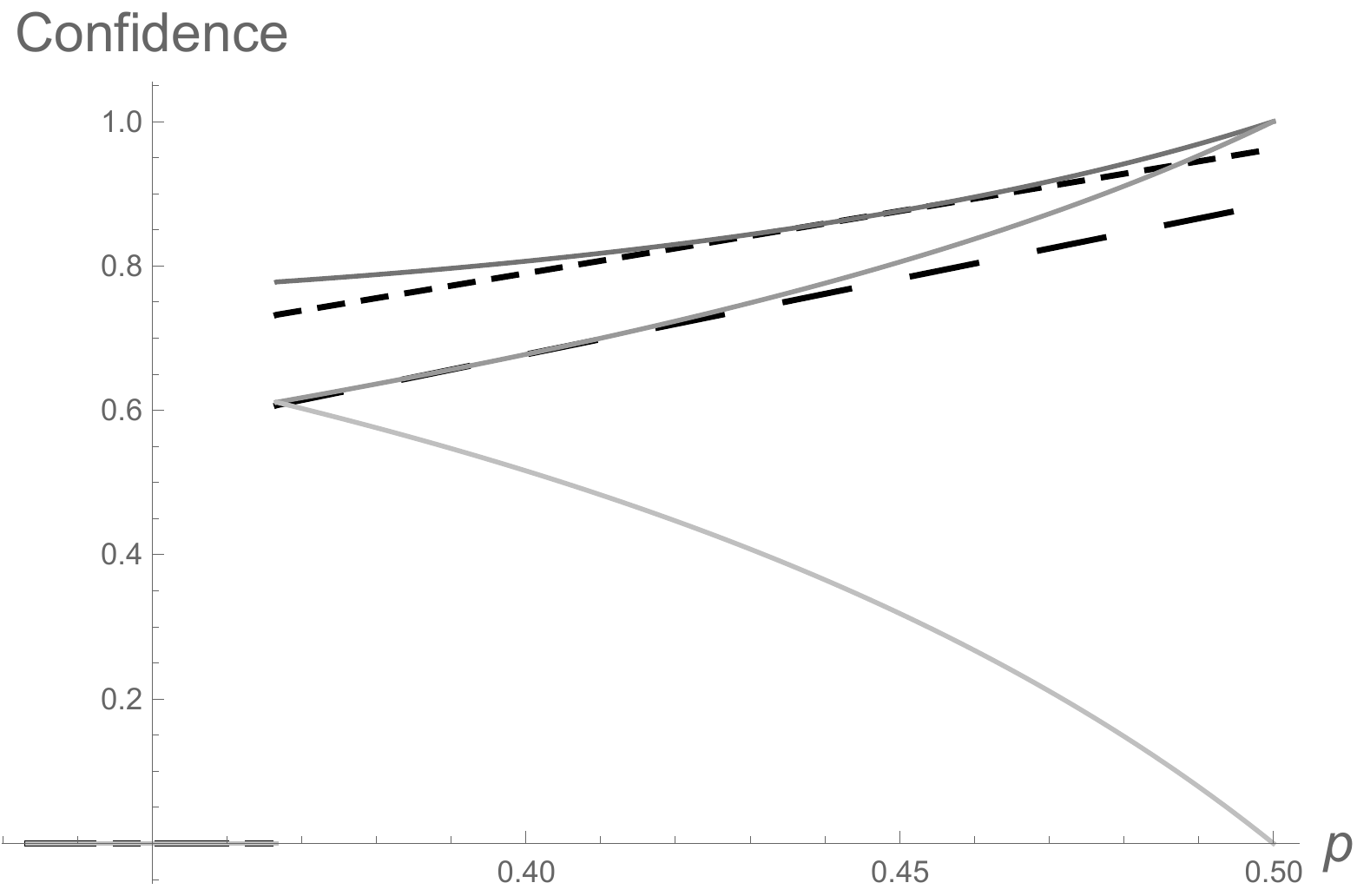}
	\caption{Graph showing the confidence in correctly identifying the signal state given the outcome of the two-element minimum-error strategy (black dotted lines) and the maximum confidence measurement for $p\in[\frac{1}{3}, \frac{1}{2}]$ and $\delta=0.1$. The lighter lines (from darkest to lightest and top to bottom) represent the MCM strategy on states $\rho_0, \rho_1$ and $\rho_2$. The higher dotted line corresponds to the minimum-error strategy on $\rho_0$, while the lower one corresponds to the same minimum-error measurement on $\rho_1$. Note that, as predicted, the most likely states are the easiest to detect in this measurement scheme. We ignore the three-outcome minimum-error measurement, as it is only optimal for a small region of the space (c.f. figure \ref{DeltaGraph}.)
		\label{MCM2}}
\end{figure}

We plot the confidence of correctly identifying each state using this measurement scheme, and compare this to the confidence using the minimum-error strategy. These can be seen in figures \ref{MCM1} and \ref{MCM2} (note that figure \ref{MCM2} only uses the two-outcome measurement, for simplicity). In both cases, the minimum-error measurement is close to optimal for $\rho_0$ and $\rho_1$. Also note how low the confidence for $\rho_2$ gets as $p$ increases - this indicates why this state is not identified in the minimum-error measurement.

\section{Conclusion}

We have investigated the optimal measurement strategies for the minimum-error and maximum confidence figures of merit for three equidistant states on the equator of the Bloch sphere with arbitrary prior probabilities, providing values for the optimal probability of correctly identifying the state in each case. The most surprising result is that, for much of the parameter space of probabilities, the optimal minimum-error measurement is a simple von Neumann measurement, and this allows optimal discrimination between these states with a minimum of resources. This is in keeping with previous results: for a completely unknown qubit state, the best measurement to estimate the state is simply a von Neumann measurement in any basis \cite{Massar95}; furthermore, the optimal intercept-resend strategy for an eavesdropper in the BB84 quantum key distribution protocol - which has four signal states - is a von Neumann measurement in the so-called Breidbart basis \cite{Breidbart}. This was also noted by Andersson \emph{et. al.}, in a case with restricted symmetry \cite{andersson2002minimum}. We have shown that the region of parameter space for which a POVM measurement is needed is rather small. This indicates that cases requiring POVM measurements are perhaps rather special, which might have implications for quantum key distribution, scalability in quantum computing, and quantum metrology. 

This paper solves the problem of optimal state discrimination between the trine states for arbitrary prior probabilities analytically; we have also shown that, for given probabilities $p_0, p_1, p_2$, there is one and only one optimal measurement - when a two-outcome measurement is optimal we know it is unique, as the measurement angle is fixed by equation (\ref{TanMment}), and, as already discussed, equation (\ref{perp}) shows that the three-element POVM must also be a unique solution. This also shows that there is no region where two- and three-outcome measurements are simultaneously optimal. This work provides a complement to that of Hunter \cite{hunter2004,hunterthesis}, which found the minimum-error strategy for arbitrary equiprobable signal states. Subsequent work presented analytical and geometric methods for arbitrary priors \cite{ha2013complete,weir2017optimal,deconinck2010qubit,bae2013structure}; what is surprising about the results presented here is the simplicity of the expressions for the optimal probability of success given in equations (\ref{P2el}) and (\ref{P3el}).

This paper also gives the maximum confidence that it is possible for a measurement to achieve on each of the trine states with arbitrary prior probabilities. This helps to identify situations in which it is sub-optimal for the minimum-error strategy to identify every signal state, as the maximum confidence possible for the least likely state tends to zero.

We hope that this work leads to new and interesting results, and we look forward to seeing other ways in which our method for tackling minimum-error discrimination problems is used.

\section{Acknowledgements}
This work was supported by the University of Glasgow College of Science and Engineering (S.C., C.H., and G.W.) and by the Royal Society Research Professorships (S.M.B., Grant No. RP150122). The authors also wish to acknowledge the invaluable help of Matthias Sonnleitner, V\'{a}clav Poto\v{c}ek and Thomas Brougham.


\end{document}